\documentclass[fleqn,10pt]{olplainarticle}
\usepackage{textcomp}
\usepackage{gensymb}
\usepackage{booktabs}
\usepackage{soul}
\usepackage{xcolor}
\usepackage{mathtools}
\RequirePackage[hidelinks]{hyperref}

\title{Regularization of ML models for Earth systems \\ by using longer model timesteps}

\author[1,2]{Raghul Parthipan}
\author[3]{Mohit Anand}
\author[4]{Hannah M Christensen}
\author[5]{Frederic Vitart}
\author[1]{Damon J Wischik}
\author[3,6]{Jakob Zscheischler}
\affil[1]{Department of Computer Science and Technology, University of Cambridge, Cambridge, UK}
\affil[2]{British Antarctic Survey, Cambridge, UK}
\affil[3]{Department of Compound Environmental Risks, Helmholtz Centre for Environmental Research -- UFZ, Leipzig, Germany}
\affil[4]{Department of Physics, University of Oxford, Oxford, UK}
\affil[5]{European Centre for Medium-Range Weather Forecasts, Reading, UK}
\affil[6]{Center for Scalable Data Analytics and Artificial Intelligence (ScaDS.AI), Leipzig/Dresden, Germany}

\correspondingauthor{Raghul Parthipan  (rp542@cam.ac.uk)}


\begin{abstract}
Regularization is a technique to improve generalization of machine learning (ML) models. 
A common form of regularization in the ML literature is to train on data where similar inputs map to different outputs. This improves generalization by preventing ML models from becoming overconfident in their predictions. This paper shows how using longer timesteps when modelling chaotic Earth systems naturally leads to more of this regularization. We show this in two domains. We explain how using longer model timesteps can improve results and demonstrate that increased regularization is one of the causes. We explain why longer model timesteps lead to improved regularization in these systems and present a procedure to pick the model timestep. We also carry out a benchmarking exercise on ORAS5 ocean reanalysis data to show that a longer model timestep (28 days) than is typically used gives realistic simulations. We suggest that there will be many opportunities to use this type of regularization in Earth system problems because the Earth system is chaotic and the regularization is so easy to implement.
\end{abstract}

\begin{document}

\flushbottom
\maketitle
\thispagestyle{empty}

\section{Introduction}



Regularization is crucial for improving the ability of models to generalize to novel domains, particularly for tasks with limited training data. Section \ref{section:regularization_back} describes how a common way to regularize models is to train on data where similar input map to different output (`SIDO') as this prevents models from becoming overconfident in their predictions. We use `similar input' to refer to input data that we believe lie close together in a latent or embedded space. In this paper, we claim that using longer timesteps for autoregressive models of Earth system data can improve generalizability and accuracy by creating more SIDO instances in the training dataset. 

Section \ref{section:results} shows that using longer model timesteps in our machine learning (ML) models improves results on an atmospheric simulator and on real ocean reanalysis data. Section \ref{section:benefits_regularization} explains \textit{how} longer model timesteps improve results by providing regularization. It begins with section \ref{section:ruling_out} addressing an alternative explanation for why longer model timesteps help: a longer timestep means a forecast can be made with fewer instances of the model ingesting its own predictions, meaning there is less chance for errors to accumulate. We provide an experiment that rules this out as the sole reason we see better results with longer model timesteps. In section \ref{section:regularization_work} we present an experiment showing that longer model timesteps specifically improve regularization. And in section \ref{section:regularization_mech}, we show that this regularization arises from how the distributions of future state $X_{t+\Delta t}$ conditioned on current states $X_t$ become more diffuse in chaotic systems as $\Delta t$ increases, which leads to more cases of SIDO. 

Section \ref{section:choosing} explains how to go about choosing the model timestep. Using too large a timestep --- meaning one beyond the predictability limit of the system --- is simply uninteresting: it is well known in the Earth system community that after a given time (the predictability limit for a particular system), just using the mean forecast (`climatology') is the best one can do. It is easy for us to calculate climatology and we do not need to train ML models to do this for us! The interesting part occurs within the predictability limit (such as 6-month lead times for the ocean). Here, a larger model timestep (such as 28 days instead of 1 day) provides added regularization, which can improve generalization. But the conditional distributions $X_{t+\Delta t}|X_t$ to be modelled become more complex as $\Delta t$ increases, which means that results may worsen if model capacity is insufficient. Therefore, for a given model capacity, there is a model timestep sweetspot between 0 and the predictability limit of the system, and we explain how to find it. 


We suspect that using longer timesteps has received limited attention in Earth system modelling because of the historical dominance of differential equation-based approaches and the historical paucity of ML tools sophisticated enough to capture complex conditional distributions. Small timesteps are typically used for traditional Earth systems models since they rely on finely resolved temporal discretization to maintain numerical stability and accuracy. Although ML allows us to set-up the task to directly model $X_{t+\Delta t}|X_t$ for large $\Delta t$, the previous generations of ML models are bad at this task since they rely on simple parametric distributions. The recent rise of generative ML models presents a new opportunity since they are better suited to capture the complex conditional distributions that arise with longer timesteps.

To underscore that larger timestep models can actually produce realistic results, in Section \ref{section:seas5_experiments}, we evaluate our model for the ocean which is trained with a model timestep of 28 days. 

Training models with larger model timesteps is an easy way to improve regularization. It creates SIDO instances without needing any user-guided data augmentation. It also does not require any changes to the underlying ML model. It would be difficult to manually augment chaotic systems' data to create SIDO instances given how sensitive these systems are to their initial conditions. For example, if we just added noise to the targets, we would likely get our models to learn the wrong relationships.

\section{Background}

This section is in two parts. The first shows how it is common to regularize ML models by training on data with SIDO instances. The second describes how regularization of any form is often used to improve simulation roll-outs in ML forecasts for weather and Earth systems. The later sections of this paper go on to show how by using a longer model timestep, we get access to SIDO regularization, which can in turn improve forecasts of the Earth system.

\subsection{Similar input -- different output (SIDO) regularization}

\label{section:regularization_back}

A single input is often compatible with multiple targets in the real world, and by training on such datasets, we can encourage our model to make appropriately uncertain predictions. Below, we cover various examples from the ML literature of how this regularization improves generalizability by reducing overconfident predictions.


Careful user intervention is often required to curate datasets with instances where it makes sense for certain inputs to map to different outputs. Although sufficiently large datasets might naturally contain such instances, smaller datasets often require deliberate data augmentation. This augmentation can be difficult to get right because the augmented additional outputs must be appropriate for the given input. If the augmentation involved just repeating the target for a given input, the model would become overconfident in its predictions. If the augmentations were more involved and meant that the patterns to be learnt became too noisy, the model may struggle to learn anything at all.



\subsubsection{Classification}

A popular technique to improve the generalization of ML classification models is to replace `hard targets' (e.g. $[1,0,0]$ for a three-class classification problem) with `soft targets' (e.g. $[0.96,0.02,0.02]$) --- created by combining hard targets with a uniform distribution over the labels. This is known as label smoothing \citep{szegedy2016rethinking}. During training, the cross-entropy loss is computed with these soft targets instead of the hard targets. This means that the network is trained to predict a smoothed version of the labels where the ground truth still has the highest probability, but other labels have a non-zero probability. Label smoothing has improved accuracy across tasks, including image classification \citep{huang2019gpipe,li2020regularization,real2019regularized,szegedy2016rethinking,zoph2018learning}, speech recognition \citep{chorowski2016towards} and in the original Transformer paper \citep{vaswani2017attention}. Similar variants include adding noise to the labels \citep{xie2016disturblabel}, and smoothing labels using a combination of a model's predicted distribution and noisy target labels \citep{reed2014training}. Label smoothing has also been used to train discriminators in GANs \citep{salimans2016improved}.

Label smoothing acts as a regularizer by preventing models from becoming overconfident in their predictions \citep{szegedy2016rethinking,zhang2017mixup}. This is because the model is encouraged to assign some non-zero probability to all classes, not just the ground truth. \citet{muller2019does} and \citet{lukasik2020does} show that this also improves the model calibration in image classification and machine translation, meaning a better alignment between prediction confidence and prediction accuracy. Moreover, \citet{lukasik2020does} show that label smoothing has a similar effect to shrinkage regularization in linear models, encouraging weights to remain close to zero. \citet{pereyra2017regularizing} show that label smoothing can be framed as equivalent to a confidence penalty which they propose. This penalty penalizes low entropy predictions i.e. where a network places all its probability on a single class, and they show that using it as a regularizer prevents these peaked distributions, leading to better generalization. 

Label smoothing involves degrading the target labels and it is unclear how much degradation is suitable. Certain amounts of smoothing can deteriorate calibration \citep{lukasik2020does}. The naive label smoothing approach of treating non-target categories as equally probable may be suboptimal --- for example, when classifying an image of a cat, the incorrect label `dog' should receive a higher probability than `chair'. Treating non-target categories as equally probable risks over-aggressive regularization in certain regions of the feature space.  To address this, \citet{zhang2021delving} propose taking into account the relationships between different categories by using information from the model's predicted label distribution. The soft labels therefore get updated during training. But this still requires balancing loss terms corresponding to hard labels and soft labels. \citet{li2020regularization} learn different smoothing strengths based on clustering the training dataset. 

\subsubsection{Regression}

Regression tasks require more care when implementing SIDO regularization due to the continuous nature of possible outputs. A straightforward extension of label smoothing involves adding noise to targets during training iterations, effectively mapping the same input to different targets at each training epoch. \citet{imani2018improving} propose a more structured approach where users first specify a target distribution (such as a Gaussian centred on the target value) and then train by minimizing the KL divergence between the model's distribution and this target distribution. Overly confident model distributions face similar penalties as in label smoothing. However, in both approaches, it is tricky to determine appropriate noise levels and distributions. If misspecified, the model will end up not learning, or learning the wrong relationships.

One example where multiple targets naturally correspond to the same input is when target data is collected using a measurement device with a known error. We know that the target we measure is not the sole target compatible with the input. The measurement uncertainty means that if we were to repeat our experiment we may see various other compatible targets. Ideally, we would repeat our experiment to provide this additional data, but this is often impractical --- for instance, in satellite measurements of the Earth. However, we can still account for this measurement uncertainty through probabilistic modelling \citep{lawrence2001estimating}. Consider a system where input data $x$ generates a true target $y$, but we only observe noisy $y'$. The joint distribution can be factorised as $p(x,y,y') = p(y'|y)p(y|x)$, where $Y'|Y=a \sim N(a, \sigma^2)$ represents Gaussian measurement error around the true $y$, and $p(y|x;\theta)$ is the learnable model distribution (which could be based on a diffusion model, for example) with learnable parameters $\theta$. Training involves maximizing the likelihood of the parameters for the observed data:
\begin{equation}
    p(y'|x) = \int p(y'|y)p(y|x;\theta) dy.
\end{equation}
This formulation shows that training the model involves adjusting $\theta$ in $p(y|x;\theta)$ for all values of $y$, for a given input $x$.  Contributions to the error are weighted by $p(y'|y)$, giving greater importance to values of $y$ closer to the observed $y'$. This approach parallels label smoothing's philosophy where many different targets are shown as compatible with the same input to prevent overconfidence, and targets closer to the ground truth contribute more to the loss function. 



\subsubsection{Generative modelling}

In generative modelling, augmentation of target images such as through rotation and additive noise is helpful for diffusion models \citep{karras2022elucidating} and GANs. Once again, care is needed with these augmentations, as naive approaches lead to the generator learning to reproduce the augmented distribution itself \citep{tran2021data,zhao2021improved,zhang2019consistency}. This is undesirable as we wish to use the augmentations as regularization but only generate clean, non-noisy, non-augmented images. To address this, DiffAugment \citep{zhao2020differentiable} augments data for both the generator and discriminator, \citet{karras2020training} propose both an augmentation mechanism which prevents leakage to the generated images and an adaptive approach to control the amount of augmentation, and \citet{zhang2024improving} use augmented real samples to help the discriminator learn that augmented samples are fake instances, so that augmentations do not leak to the generator.

\subsubsection{Multitask learning}

Multitask learning \citep{caruana1997multitask,crawshaw2020multi,ruder2017overview} is another setting where better generalization is achieved by training on data where the same input is mapped to different tasks/outputs. It has successfully been used across ML from natural language processing \citep{bingel2017identifying,collobert2008unified,liu2015representation,luong2015multi} to computer vision \citep{liu2019end,girshick2015fast}.  In the weather domain, AtmoRep's \citep{lessig2023atmorep} self-supervised masked training objective can be interpreted as an implicit form of multitask learning --- the patch masking technique inherently allows multiple outputs to correspond to a single input. 
Multitask learning helps generalization because the model learns data representations which are compatible with the different outputs. \citet{caruana1997multitask} describes the benefits as a form of implicit data augmentation, noting that since all tasks have some noise, training a model on different tasks with differing noise patterns helps it learn more general patterns. Again, it is both important and non-trivial to determine which other tasks/targets the model should be trained on \citep{caruana1997multitask, crawshaw2020multi}. A poor choice of additional tasks may lead to worsened model performance. 


\subsection{Regularization to improve roll-outs in ML weather forecasts}

Simulation roll-outs often test a model’s ability to generalize by exposing it to scenarios beyond its training distribution. Many models make forecasts iteratively: they ingest an initial state, generate an output, feed that output back in, and repeat the process. Small errors accumulate at each step (error accumulation), pushing the model further from its training distribution. This can trigger a negative feedback loop with the model performing even worse at subsequent steps, leading to degradation of forecasts. Error accumulation during roll-outs is well-documented in  ML weather forecasting  \citep{bi2023accurate,chen2023fuxi,nguyen2023scaling,lam2022graphcast}, ML parameterization work \citep{brenowitz2018prognostic,brenowitz2019spatially} and dynamical systems work \citep{balogh2021toy,liu2020hierarchical,sanchez2020learning}. The same challenge arises in sequence modelling more broadly  \citep{bengio_scheduled_sampling,schmidt2019generalization,leblond2017searnn,ranzato2015sequence,lamb2016professor}. 

Improved regularization can improve roll-out performance by allowing models to cope better when they ingest data (often self-generated) out of the training distribution \citep{schmidt2019generalization}. A classic way to regularize is to train a model on more ground truth data. \citet{balogh2021toy} showed that training ML models on data from regions outside a system’s attractor can stabilize roll-outs. If a model encounters such data at test time, it has already learned how to respond. Similarly, \citet{rasp2020} trained an ML model by running it alongside a high-resolution (HR) simulation treated as the ground truth. The HR was nudged to align with the ML model, and its outputs were used as training targets. In this way, the ML model is supposed to learn how to behave similarly to how the ground truth model would behave if encountering a similar (potentially error-containing) state. This approach improves stability but is computationally expensive and impractical for many real-world systems where no ground truth simulator can be easily run or is even available.

A popular alternative is training models on their own generated data. This acts as a form of regularization because the model is shown additional, synthetic (i.e. generated) data during training. Scheduled Sampling \citep{bengio_scheduled_sampling} introduced this concept, and there are numerous variants of it based on multi-step loss functions such as the pushforward trick \citep{brandstetter2022message}. These methods involve generating trajectories from the ML model and then use a loss function to penalize the difference between the generated trajectories and the true trajectories. This roll-out training is used in ML weather prediction models such as GraphCast \citep{lam2022graphcast}, FourCastNet \citep{pathak2022fourcastnet}, Stormer \citep{nguyen2023scaling}, FuXi \citep{chen2023fuxi} and Aurora \citep{bodnar2024aurora}. 


The GenCast \citep{price2023gencast} model for weather forecasting is notable in how it produces stable trajectories without requiring roll-out training. Two factors may explain this: (1) diffusion models like GenCast can capture conditional spatial dependencies since they are generative models, unlike architectures such as GraphCast \citep{lam2022graphcast} which assume spatial independence conditional on prior states. This may translate to improved generalizability. (2) Diffusion models may have a natural tendency to be underdispersive, reducing the chance of sampling from distribution tails. This implicitly achieves a similar effect to top-k/top-p sampling in language models, where improbable samples are suppressed. Such techniques tackle error accumulation by making the model underdispersive at inference time, reducing the chances of the model having to ingest data not seen during training. While this property may help mitigate error accumulation during simulations, it raises a potential concern: reducing the likelihood of tail events may inadvertently suppress extreme weather events, which are often critical in weather forecasting.

Regularization can also be imposed via loss function penalties as is standard in ML. \citet{bretherton2022correcting} applied L2 regularization to improve roll-out stability, but found it degraded one-step-ahead accuracy, suggesting overly strong penalization. Physical constraints are another option: \citet{yuval2021} enforced energy conservation in an ML parameterization model, while \citet{brenowitz2019spatially} adjusted input processing to prevent models from learning unphysical correlations.

\section{Can a longer model timestep be helpful?}
\label{section:results}

Before jumping into \textit{why} we believe a longer timestep provides added regularization and \textit{how to decide} on the timestep, we first demonstrate that a longer timestep can indeed help. We do this not only to convince the reader that the model timestep is a worthwhile design choice to explore when building models, but also to help the reader develop intuition about how the model timestep affects forecast performance. 

We present two applications in this section. The first uses simulated data from the Lorenz 96 system. The second uses real ocean reanalysis data (ORAS5). In both cases, the task is to forecast the evolution of the system starting from a given initial state. Our experiments involve training models with different model timesteps and examining how forecast performance varies with the model timestep. For each application, we provide details on the dataset, give a brief overview of the models we train and then present the experimental results.

\subsection{Lorenz 96}

\subsubsection{Dataset}

The Lorenz 96 system is a toy system for atmospheric circulation that was developed by Lorenz to understand how atmospheric processes of different scales interact \citep{lorenz1996predictability}. It is extensively used in the `parameterization' field in weather and climate to study how small-scale processes in the atmosphere influence larger-scale ones \citep{hannah_bespoke,crommelin2008subgrid,gan_hannah,kwasniok2012data,rasp2020}. The system we use (following the specification in \citet{gan_hannah}) generates trajectories of variables on two scales: coarse and fine. Our task is to model the evolution of the 8 coarse variables. Note that one unit of model time is considered equivalent to 5 atmospheric days in the sense that this is how long it takes for forecast errors to double in each respective system. For reference, \citet{gan_hannah} and \citet{hannah_bespoke} used a timestep of 0.005 model time units when modelling the system. Details of the train/test split are provided in Appendix \ref{appendix:l96}. 

\subsubsection{Model}

We model the system by first quantizing the data, and then learning a categorical distribution over the resulting discrete data. Concretely, we quantize the state $X_t \in \mathbb{R}^8$ using 100 uniform bins to get $Z_t \in \{1,2,...,100\}^8$. The next discrete state, $Z_{t+\Delta t}$, is then sampled from the categorical distribution $Z_{t+\Delta t} \sim Cat\big(p_\theta(Z_t)\big)$, where $p_\theta$ is a neural network with a single hidden layer of dimension 32. Finally, the generated sequence is dequantized by mapping bins to their bin centres on the real line. This quantization approach for modelling continuous data is a simple way to capture sophisticated distributions \citep{ansari2024chronos,sonderby2020metnet}.

\subsubsection{Experiments}

\begin{figure}[h!]
\centering
\includegraphics[width=0.75\linewidth]{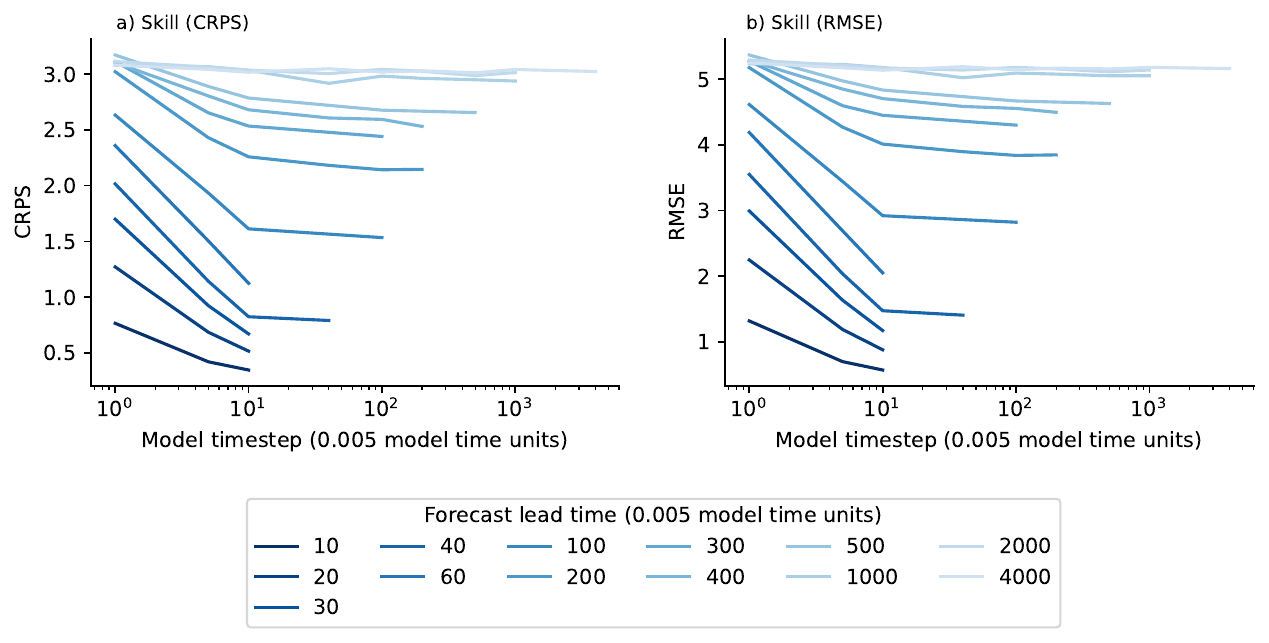}
\caption{Effect of model timestep on Lorenz 96 skill scores. This compares the skill (y-axis) in terms of CRPS ((a)) and RMSE ((b)) as a function of timestep (x-axis) of nine separate models. Lines join model skill scores for a particular forecast lead time. The darkest line is for the $10 \times 0.005 $ model time unit lead time, and the lightest line is for the $4000 \times 0.005$ model time unit lead time. In all cases, lower is better.}
\label{fig:l96_skill}
\end{figure}

\begin{figure}[h!]
\centering
\includegraphics[width=0.5\linewidth]{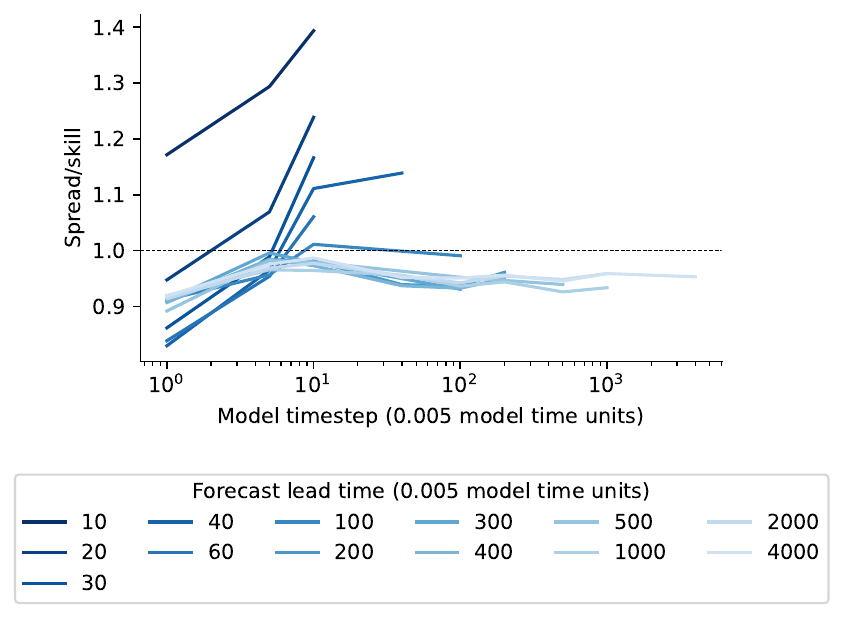}
\caption{Effect of model timestep on Lorenz 96 spread/skill scores. This compares the spread/skill (y-axis) as a function of timestep (x-axis) of nine separate models. Lines join model skill scores for a particular forecast lead time. The darkest line is for the $10 \times 0.005 $ model time unit lead time, and the lightest line is for the $4000 \times 0.005$ model time unit lead time. In all cases, a score closer to one is better.}
\label{fig:l96_spread_skill}
\end{figure}

We created 20-member ensemble forecasts from 400 initial conditions. Figure \ref{fig:l96_skill} and Figure \ref{fig:l96_spread_skill} show how the forecast skill and forecast calibration vary with model timesteps of $1 \times 0.005$, $5 \times 0.005$, $10 \times 0.005$, $40 \times 0.005$, $100 \times 0.005$, $200 \times 0.005$, $500 \times 0.005$, $1000 \times 0.005$ and $4000 \times 0.005$ model time units. In Figure \ref{fig:l96_skill}, the x-axis is the model timestep and the y-axis is the skill, which we describe in the next paragraph. Lower skill values are better. Lines join forecasts at a particular lead time, allowing for comparison of how a change in model timestep affects the forecast skill at that lead time. The lines do not all cover the full x-axis, because not all the models make forecasts at every lead time. For example, at a lead time of $10 \times 0.005$ model time units (darkest blue line), only the models with timesteps of $1 \times 0.005$, $5 \times 0.005$ and $10 \times 0.005$ model time units can make that forecast. This is also the case for the lead time of $60 \times 0.005$ model time units. 

The forecast skill improves as the model timestep increases within the predictable lead times (roughly up to $1000 \times 0.005$ model time units for the Lorenz 96). At longer lead times, the skill does not notably improve with the model timestep. This is because predictability will be lost and the best that can be done is to output climatology. Although we trained models with timesteps longer than $1000 \times 0.005$ model time units for this experiment, it would be wasteful to do this in an operational setting when we know this is beyond the predictable timescale.



We measure forecast skill in this paper using the Continuous Ranked Probability Score (CRPS) and the Root Mean Squared Error (RMSE). The CRPS is a standard measure of the skill of a probabilistic forecast. It measures how well the marginal distributions of a forecast represent the ground truth, and it is minimized, in expectation, by a forecast whose marginals reflect true predictive uncertainty. The RMSE measures how the mean of an ensemble forecast matches ground truth. It does not account for uncertainty, but it is a central metric in deterministic forecasts. See Appendix \ref{appendix:evaluation_details} for their definitions.

Figure \ref{fig:l96_spread_skill} can be interpreted similarly to Figure \ref{fig:l96_skill} but with the y-axis now the spread/skill score. Forecast calibration (i.e. a degree of uncertainty which matches the size of errors \citep{fortin2014should}) can be measured using the spread/skill score, defined in Appendix \ref{appendix:evaluation_details}. This ratio should be 1 for a perfect ensemble forecast, with values greater than 1 suggestive of over-dispersion (an under-confident forecast), and values less than 1 suggestive of under-dispersion (over-confidence). From Figure \ref{fig:l96_spread_skill}, we see that the spread/skill scores vary widely between underdispersion and overdispersion for the smaller model timesteps. At larger timesteps the scores stabilise around 0.95.

\subsection{ORAS5 ocean reanalysis}

\subsubsection{Dataset}

Forecasting the atmosphere out to the 6-month horizon (long-range) requires accurate tracking of the ocean, especially sea surface temperatures (SSTs). Long-range forecasts depend on the predictability offered by boundary conditions, such as SSTs and ice surfaces in contact with the atmosphere. Therefore, it is important to get the initial condition and subsequent evolution of the ocean right.

We use the ECMWF's ORAS5 dataset \citep{zuo2019ecmwf} which is a global ocean reanalysis product, meaning it combines observations of the oceans with reforecasts of the ocean evolution to obtain a `best estimate' of the ocean's state. ORAS5 is used to initialize the ECMWF's long-range forecasting model, SEAS5 \citep{johnson2019seas5}. We use a version of ORAS5 which is stored at a $1\degree$ latitude/longitude resolution. Our model is trained on 25 years of ORAS5 \citep{zuo2019ecmwf} from 1993 to 2018, and further details are provided in Appendix \ref{appendix:oras5}.

Our task is to model 7 upper ocean variables across the global ocean, including SST (see Table \ref{tab:variables_used} for details). SST evolution is particularly important for long-range prediction as SSTs retain information and exert influence over longer timescales. This contrasts with small-scale atmospheric features which are crucial for short- to medium-range forecasting, but have a diminishing influence over time due to the chaotic nature of weather systems. For example, the El Niño/Southern Oscillation (ENSO) is an important recurring climate pattern involving changes in the temperature of waters in the central and eastern tropical Pacific Ocean. ENSO cycles between warmer (El Niño) and cooler (La Niña) SSTs, leading to characteristic changes across vast regions of the globe, with impacts lasting from a few months to several years. SSTs also influence local climates, especially in coastal areas. For instance, warmer SSTs increase the likelihood of tropical cyclone formation and can intensify existing storms by providing additional energy, while cooler SSTs can inhibit their development and strengthening.

\begin{figure}[]
\centering
\includegraphics[width=0.9\linewidth]{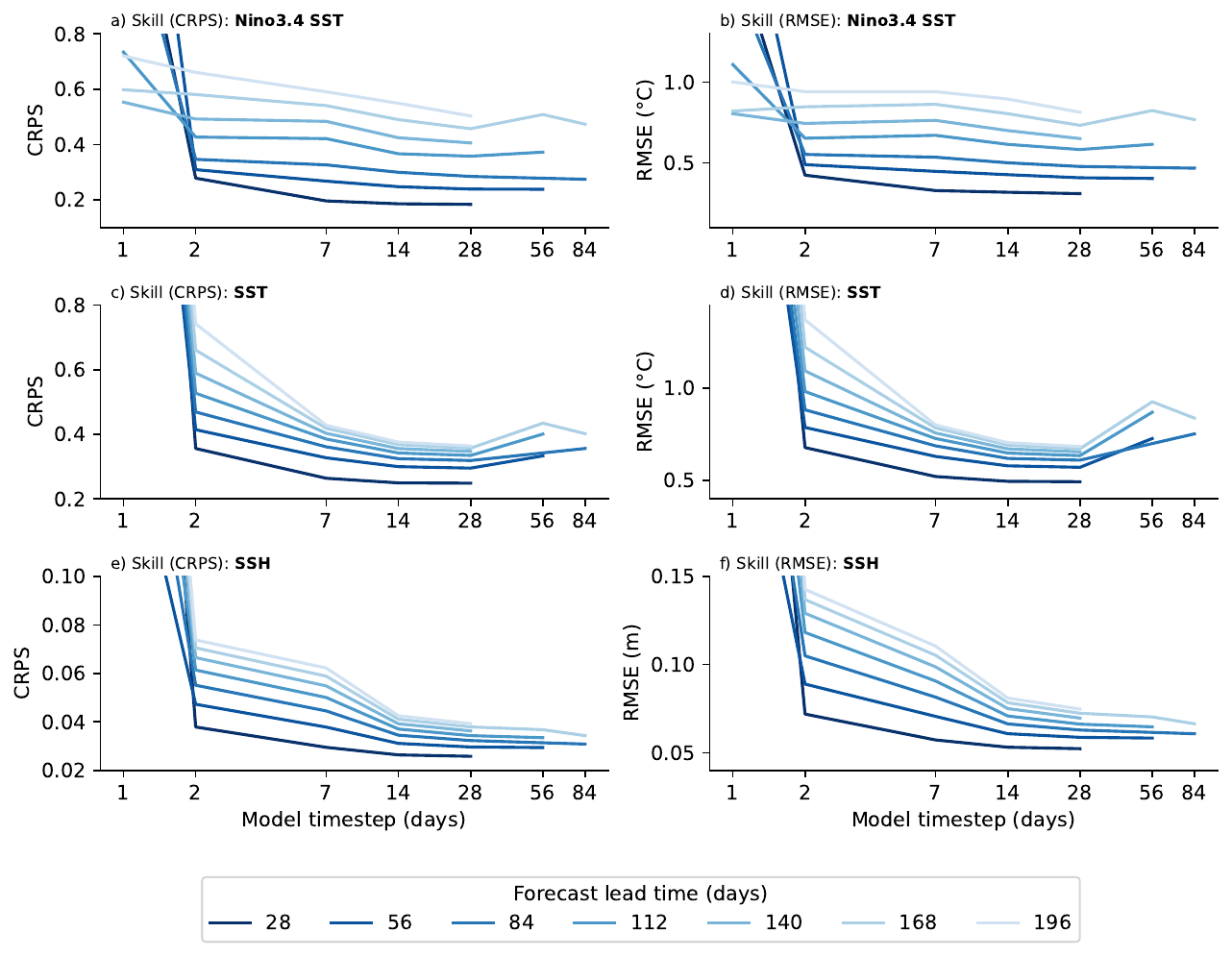}
\caption{Effect of model timestep on ORAS5 skill scores for 2019-2022. This compares the skill (y-axis) in terms of CRPS ((a), (c), (e)) and RMSE ((b), (d), (f)) as a function of timestep (x-axis) of seven separate models (timesteps 1, 2, 7, 14, 28, 56 and 84 days).  Skill is evaluated for Nino3.4 SST, whole-ocean SST, and whole-ocean SSH. Lines join model skill scores for a particular forecast lead time. The darkest line is for the 28-day lead time, and the lightest line is for the 196-day lead time. Skill  scores for the 84-timestep model are only available at lead times of 84 and 168. In all cases, lower is better. The x-axis is logarithmic.}
\label{fig:model_timestep_comparison}
\end{figure}

\subsubsection{Model}

We introduce our probabilistic model for the ocean --- SeaDiffusion --- which models the probability distribution, $p(X_{t+\Delta t}|X_t)$, of a future ocean state $X_{t+\Delta t}$ conditioned on the current state. A forecast trajectory, $X_{t: K \Delta t}$, of length $K$ steps, is modelled by conditioning on the initial state $X_0$ and factoring the joint distribution as so,
\begin{equation}
    p(X_{t: K \Delta t}|X_0) = \prod_{t=0}^{(K-1)\Delta t} p(X_{t + \Delta t} | X_t),
\end{equation}
where each state is sampled autoregressively.

The representation of the $1 \degree$ global ocean state, $X$, is composed of sea surface height (SSH), and temperature at 6 depth levels in the upper ocean, including SST. A single variable in the ocean state is represented by a 112 x 240 latitude/longitude grid. An input state comprising $I$ channels' worth of 112 x 240 gridded data is taken in, and samples of shape 7 x 112 x 240 are generated (see Table \ref{tab:variables_used} for details).

SeaDiffusion is a conditional diffusion model \citep{karras2022elucidating,ho2020denoising,song2019generative,song2020score,sohl2015deep}, which is a type of generative ML model capable of sampling from sophisticated data distributions. Diffusion models work by iteratively removing noise from a candidate state. A future ocean state, $X_{t+\Delta t}$
is produced by iteratively refining an initial state, $Z_{t+\Delta t}^0 \sim N(0,I)$, conditioned
on the previous state $X_t$. A different generative sample can be created by initializing the process with a different noise sample (i.e. $Z_{t+\Delta t}^0$), which allows an ensemble of trajectories to be made. We use the diffusion framework presented by \citet{karras2022elucidating}. See Appendix \ref{appendix:seadiffusion}
for further details.

At each stage of the iterative refinement process, SeaDiffusion uses a modified version of the ADM U-net architecture \citep{dhariwal2021diffusion,karras2024analyzing}, which combines a U-net architecture \citep{ronneberger2015u}  with self-attention layers \citep{vaswani2017attention}. Its variants have been widely adopted in large-scale diffusion models, including Stable Diffusion \citep{rombach2021high}, DALL-E 2 \citep{nichol2021glide, ramesh2022hierarchical}, and Imagen \citep{saharia2022photorealistic}.

\subsubsection{Experiments}

We created 51-member forecasts from 48 initial conditions (start of every month in the 2019-2022 test set). We evaluate forecast skill and calibration for SSH, SST, and Nino3.4 SST which is the average SST in the Nino3.4 region (5°N–5°S, 170°W–120°W) in the central equatorial Pacific. The region is important for monitoring ENSO phases. We use the same metrics as for the Lorenz 96 in the previous section. Figure \ref{fig:model_timestep_comparison} shows how skill scores vary with the SeaDiffusion model timesteps. The skill scores in all cases improve from 1 to 28-day model timesteps. In Figure \ref{fig:model_timestep_comparison}a, b, e and f, the skill scores improve further to the 84-day model, whereas in c and d, they worsen. This worsening at longer timesteps is likely due to limited model capacity meaning that the more complex conditional distributions for the 56-day and 84-day model are not properly learnt. We note that all the forecast lead times we analyse are within the predictable timescales for the ocean.

\begin{figure}[]
\centering
\includegraphics[width=0.65\linewidth]{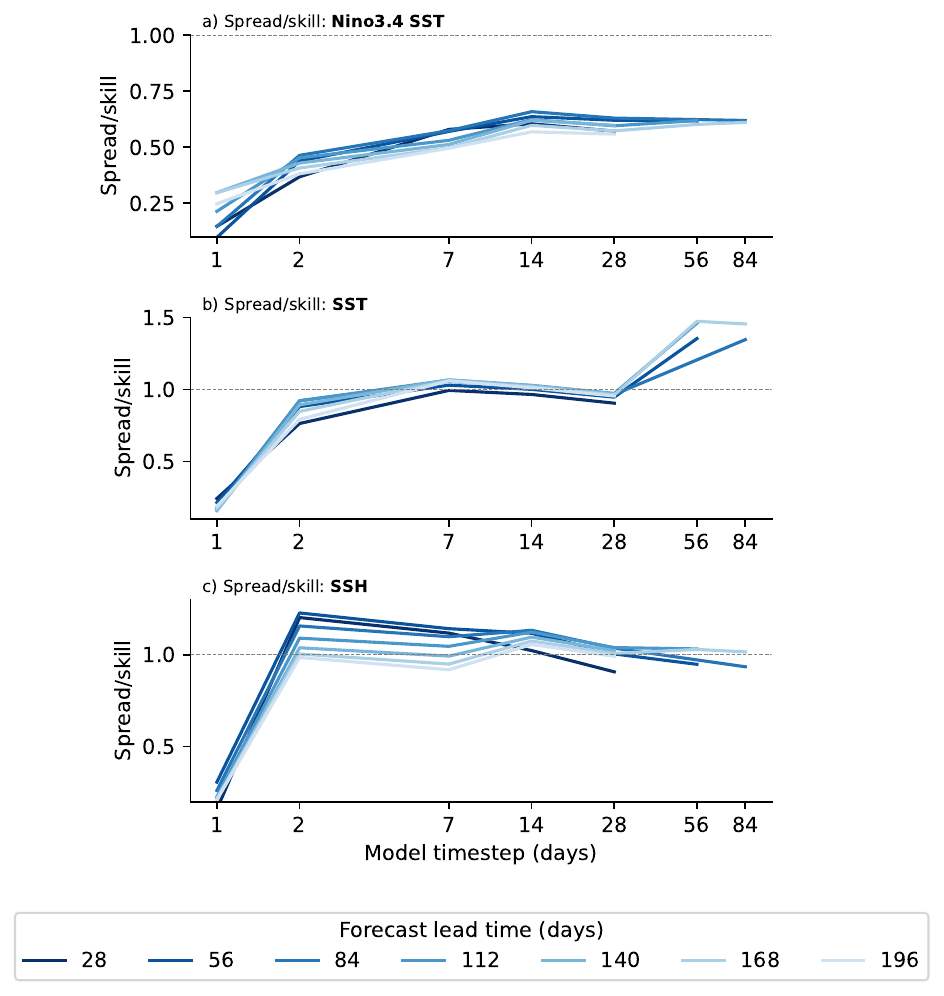}
\caption{Effect of model timestep on ORAS5 spread/skill scores for 2019-2022. This compares the spread/skill (y-axis) as a function of timestep (x-axis) of seven separate models (timesteps 1, 2, 7, 14, 28, 56 and 84 days).  Spread/skill is evaluated for Nino3.4 SST, whole-ocean SST, and whole-ocean SSH. Lines join model skill scores for a particular forecast lead time. The darkest line is for the 28-day lead time, and the lightest line is for the 196-day lead time. Skill scores for the 84-timestep model are only available at lead times of 84 and 168. In all cases, a score closer to one is better. The x-axis is logarithmic.}
\label{fig:model_timestep_comparison_ss}
\end{figure}

Figure \ref{fig:model_timestep_comparison_ss} shows how spread/skill varies with the different SeaDiffusion models. Figure \ref{fig:model_timestep_comparison_ss}a shows the spread/skill is underdispersive in all cases, but improves for longer model timesteps. For the global SST in Figure \ref{fig:model_timestep_comparison_ss}b, the spread/skill is best at 7, 14 and 28 days, becoming overdispersive beyond. And for global SSH, in Figure \ref{fig:model_timestep_comparison_ss}c, the spread/skill again tends to improve for longer model timesteps and is closest to 1 for the 28-day model.



\section{Are the benefits of longer timesteps due to better regularization?}
\label{section:benefits_regularization}

Why is it that we get better performance with longer timesteps? Is it because of the regularization which we claim is taking place, or is it solely due to something less interesting. And, if regularization is taking place, what enables this? This section addresses these questions.

\subsection{Ruling out obvious explanations}
\label{section:ruling_out}

The obvious explanation for why longer timesteps improve forecast performance is that the forecast model needs to be iterated fewer times. This means that there is less chance for model-induced roll-out errors because the model is simply rolled-out fewer times. This is why Pangu-Weather \citep{bi2023accurate} and \citet{liu2020hierarchical} use longer timesteps to address error accumulation. Another obvious possibility is network capacity. Within the predictable timescales, it can be more difficult to learn the conditional distributions $p(X_{t+\Delta t}|X_t)$ for larger $\Delta t$ as these will become more complex. Therefore, a given model capacity may lead to overfitting when learning $p(X_{t+1}|X_t)$, but not end up with overfitting when learning $p(X_{t+28}|X_t)$ simply because the latter is a harder task.

To address the first point, we propose an experiment to show that a longer timestep is beneficial even when the total number of roll-out iterations is controlled for. If the benefits seen in Section \ref{section:results} are just due to different numbers of model iterations, we would expect to see similar degradation in forecasts when this is controlled for. Our experiment is as follows: take a SeaDiffusion model with a given model timestep initialized at 1st January 2019. Run an ensemble of 51 simulations forward for 4700 iterations. Note, this is possible as the only non-prognostic variables which SeaDiffusion ingests are clock features and static features (latitude and longitude) which we know for the full length of the simulation. Plot the minimum SST and maximum SST for each of these simulations over the length of the roll-out. If these leave the expected range, this indicates the model is unstable. Repeat this for the SeaDiffusion models with different timesteps.

\begin{figure}[h!]
\centering
\includegraphics[width=1.1\linewidth, trim= 0 150 0 80, clip]{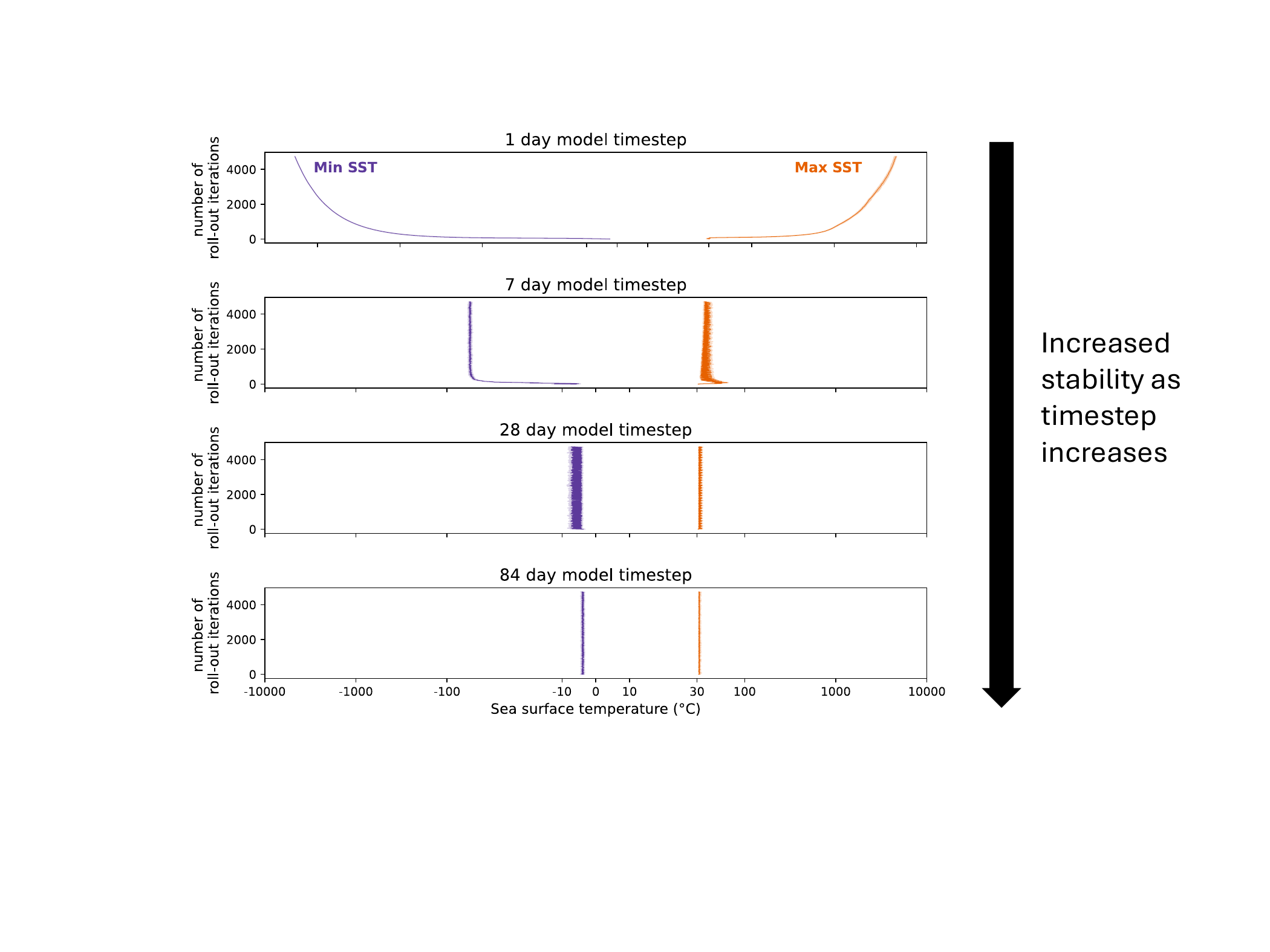}
\caption{Effect of model timestep on long-run stability. For each model, a 51-member ensemble was run out for 4700 roll-out iterations (y-axis). At each iteration, the median and interquartile range of the minimum sea surface temperature (SST) across ensemble members is plotted in purple. Similarly, the median and interquartile range of the maximum SST across ensemble members is plotted in orange. The SST x-axis is logarithmic.}
\label{fig:sst_stability}
\end{figure}

We conduct this experiment and show the results in Figure \ref{fig:sst_stability}. The models with a larger timestep are more stable throughout the simulation period. The 28 day model and the 84 day model maintain minimum and maximum SSTs which are close to their expected ranges. The 7-day model's minimum SST for all ensembles deviates to around -65 $\degree C$ before stabilising, whereas the 1-day model quickly becomes unstable. The 28-day and 84-day models are therefore stable for 360 years and 1081 years, respectively. We conclude that the benefits from using a longer timestep are not just down to running the model for fewer iterations.

We address the network capacity explanation in our experiment in the next section.

\subsection{Regularization is at work}
\label{section:regularization_work}
\begin{figure}[h!]
\centering
\includegraphics[width=0.5\linewidth]{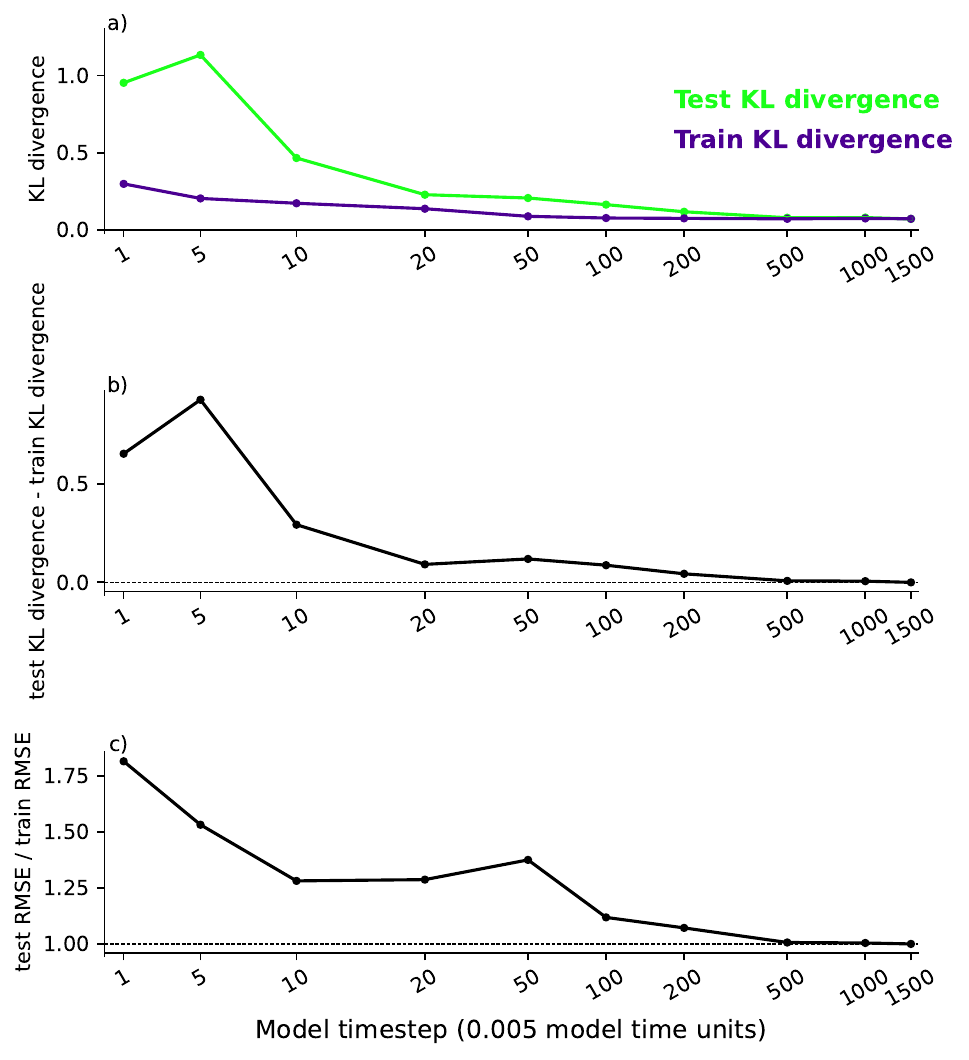}
\caption{Regularization benefits of using a longer model timestep $\Delta t$ in the Lorenz 96 system. a) KL divergence between the modeled and true conditional distributions of $X_{t+\Delta t}^i|X_t^j$ on the test set (green) and train set (purple) as a function of model timestep. Both decrease with increasing $\Delta t$, indicating improved modelling of the conditional distributions and improved generalization. Here, i and j index the 8 Lorenz 96 variables. b) Difference between test and train KL divergence, showing that overfitting is more pronounced for small timesteps and diminishes as $\Delta t$ increases. c) Ratio of test RMSE to train RMSE. This ratio approaches 1 for larger timesteps indicating improved generalization.}
\label{fig:l96_regularization}
\end{figure}

We can directly show the benefits of a longer timestep in how it improves regularization. The success of a regularization technique is often measured by how it improves test set performance as a proxy for how it improves generalization. The naive thing for us to do would then be to compare the skill of the next-step forecast of the state $X_{t+\Delta t}$ given $X_t$, for different values of model timestep $\Delta t$. This would not work as it involves comparing errors of entirely different quantities: the error when modelling $X_{t+100}|X_{t}$ is bound to be larger than $X_{t+1}|X_{t}$ because chaotic systems become less predictable as the time horizon increases. 

We devise the following experiment on the Lorenz 96 to measure generalizability in a comparable way based on measuring the KL divergence between the modelled distributions and the truth. A smaller KL divergence indicates two distributions are better matched, with 0 being the smallest possible value. If longer timesteps improve regularization, then $\textrm{KL}(p_{\textrm{truth}}(X_{t+\Delta t}^i|X_t^j) \, || \, p_{\textrm{model}}(X_{t+\Delta t}^i|X_t^j))$ should decrease for larger model timesteps as the modelled distributions of the conditional $X_{t+\Delta t}^i|X_t^j$ become closer to the truth. Note that here $i$ and $j$ denote one of the 8 Lorenz 96 variables. The experiment is as follows: create a test set for all data where either $X_t^0$ or $X_t^1$ or $X_t^2 > 14$. Keep a validation set for data where $X_t^3 > 14$, and when training the models save the ones which do best on the validation set. Now, create 100 uniform bins, $b = [1,...,100]$ for each of the 8 $X_t$ variables. For the test set, start by creating the density histogram for $X_{t+\Delta t}^0|X_t^0 \in \textrm{bin}_1$ for both the actual data and for our model, based on doing conditional sampling from the model. Note that each model is \textit{only used to sample one model timestep ahead} i.e. next step prediction. There are no multi-step rollouts here. Measure the KL divergence between these true and modelled density histograms. This can be done analytically. Next, compute the average of the KL divergences over $X_t^0 \in \textrm{bin}_b$ for $b = [1,...,100]$, and also over all $X_{t+\Delta t}^i$ and $X_{t}^j$ for all 64 $i$ and $j$ combinations. Repeat this for the different $\Delta t$. 

Figure \ref{fig:l96_regularization} shows the results of this experiment. Figure \ref{fig:l96_regularization}a) shows in green how the KL divergence on the test set between the truth and the model decreases as the model timestep increases. In purple we also plot the train set KL divergence, which also decreases as the model timestep increases. The train KL divergence improving with larger model timesteps shows that for this experiment, there is no limitation in model capacity here: the conditional distributions for the larger timesteps are modelled as well if not better than the smaller ones on the train set. The improvement in test set KL divergence is therefore not simply due to a restriction in model capacity. We conclude that the longer timestep is indeed improving regularization.

Figure \ref{fig:l96_regularization}b) is derived from a) and clearly shows how the difference between the train and test KL divergence decreases as the model timestep increases. For smaller timesteps the models are overfitting, whereas their performance on the test and train set becomes more similar for the larger timesteps. Figure \ref{fig:l96_regularization}c) shows the ratio of the test set RMSE to the train set RMSE. We take the ratio to allow for a meaningful comparison between the errors for different forecast lead times. Again, we see this drops closer to one indicating better match between train and test set performance. 


\subsection{Regularization mechanism}
\label{section:regularization_mech}

Knowing that the performance improvement is due to regularization is different to knowing \textit{how} that regularization comes about. As has been detailed extensively in the existing literature (Section \ref{section:regularization_back}), having training data with SIDO acts as a regularizer, by reducing the model's chances of being overconfident when making predictions. We propose that a longer timestep is what allows access to more SIDO cases in the training dataset. This subsection examines why we would expect these SIDO cases to appear when longer model timesteps are used, and provide illustrative examples.

We expect chaotic systems to show more cases of SIDO at larger timesteps due to sensitivity to initial conditions. Chaotic systems, like the Earth system, are characterized by extreme sensitivity to initial conditions, as first identified by \citet{lorenz1963deterministic}. This means that nearby points in the system's phase space separate at an exponential rate. For times far enough into the future this means that prediction becomes practically impossible. But in the interim, there is both predictability and instances of SIDO.  

To illustrate SIDO for a 1D system is easy --- you could plot the 2D histogram of $X_{t+\Delta t}$ vs $X_t$ and examine how this changes with $\Delta t$. This is not possible with our systems due to their high dimensionality. Therefore, we first embed the high dimensional states to 1D and then visualize them using 2D histogram. We use TSNE for the visualizations here. TSNE is a nonlinear dimensionality reduction technique which maps high-dimensional data to a lower-dimensional space while preserving local similarities. But what we will see is not simply an artefact of having used TSNE. In the Appendix we also include similar examples having used Principal Component Analysis and having just taken the mean of a state to compress it to 1D (Figure \ref{fig:pca_mean_plots}).

\begin{figure}[h!]
\centering
\includegraphics[width = 0.8\linewidth]{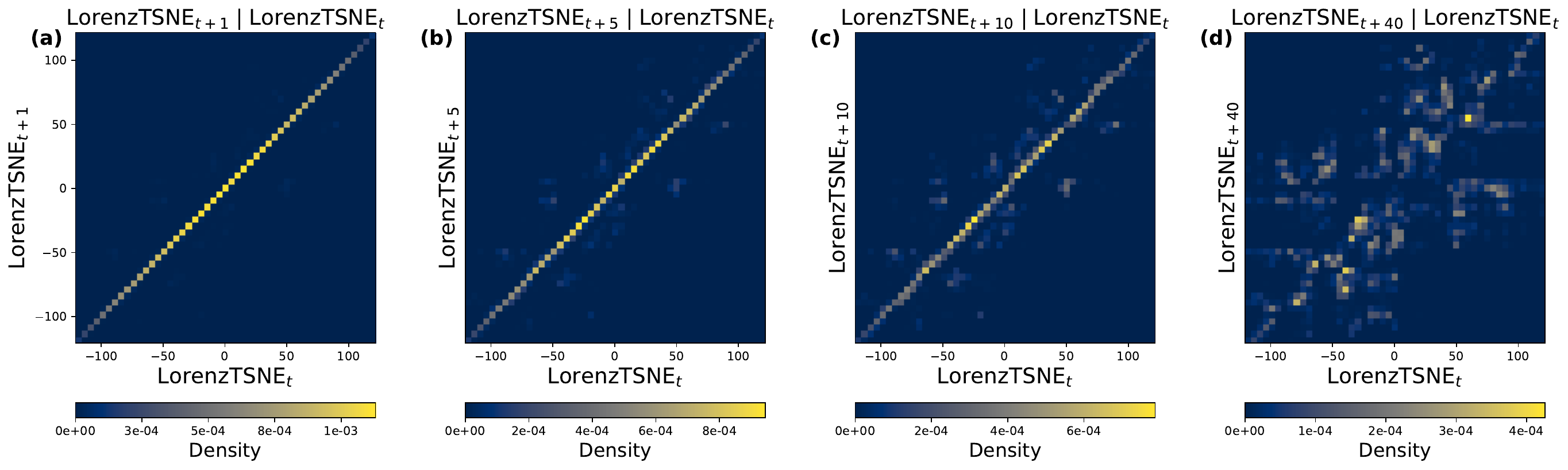}
\caption{2D histograms of TSNE embeddings of Lorenz 96 reanalysis, approximating joint distributions. LorenzTSNE are 1D TSNE embeddings of the 8D Lorenz 96 variables. a), b), c) and d) are for timesteps of 1 x 0.005, 5 x 0.005, 10 x 0.005 and 40 x 0.005 model time units respectively.}
\label{fig:lorenz_tsne}
\end{figure}

\begin{figure}[h!]
\centering
\includegraphics[width = 0.9\linewidth, trim = 160 20 150 40, clip]{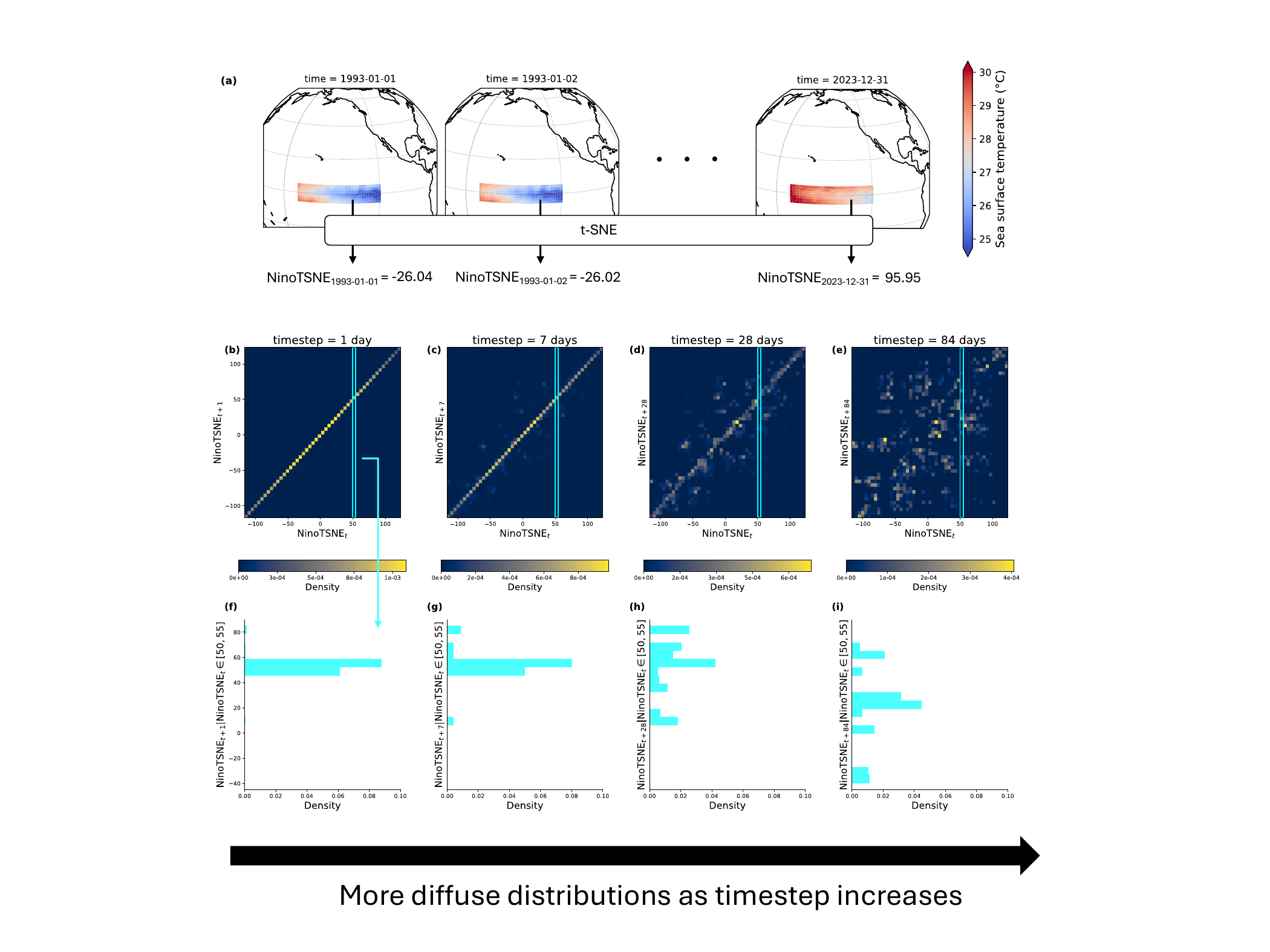}
\caption{2D histograms of TSNE embeddings of ORAS5 reanalysis, approximating joint distributions. NinoTSNE are 1D TSNE embeddings of the 238D sea surface temperature (SST) in the Nino3.4 region. a), b), c) and d) are for timesteps of 1, 7, 28 and 84 days respectively. Conditional distributions of $\text{NinoTSNE}_{t+\delta t} | \text{NinoTSNE}_{t} $ are approximated by vertical slices in the 2D histogram.}
\label{fig:nino_tsne}
\end{figure}

In Figure \ref{fig:lorenz_tsne}, we see the TSNE histograms for the Lorenz 96. In Figure \ref{fig:lorenz_tsne}a), similar input states, LorenzTSNE$_t$ lead to similar output states at $t+1$, LorenzTSNE$_{t+1}$. But as the timestep increases from a) to d), we see that similar input states correspond to vastly different outputs. In Figure \ref{fig:nino_tsne}, we see a similar example for the SST in the Nino3.4 region. Figure \ref{fig:nino_tsne}a) shows how the TSNE is performed: SST from the Nino3.4 region (238D tensor) for every day in the training set is projected using TSNE to 1D. These are then plotted in b) to e). Again, we see that a larger timestep results in more cases of similar inputs leading to different outputs. In f) to i) we plot conditional distributions of $\textrm{NinoTSNE}_{t+\Delta t}|\textrm{NinoTSNE}_t \in [50,55]$, where $\textrm{NinoTSNE}_t \in [50,55]$ is represented by the blue rectangle in plots b) to e). These conditional distributions are what our models are trying to learn, and we see that they become more diffuse as the timestep increases.

\section{How should you choose the model timestep?}
\label{section:choosing}

Section \ref{section:results} showed the benefits from a longer model timestep. Section \ref{section:benefits_regularization} showed part of these benefits are due to the regularization that accompanies the choice of a longer model timestep in chaotic systems. This section examines how you should you go about choosing the model timestep in practical model building cases. 


There are constraints that prevent us from simply choosing an arbitrarily long model timestep. First, I) the timescale of predictability sets an upper bound on the model timestep. As we have noted, just predicting climatology is the best that can be done beyond these timescales. There is no need for an ML model beyond these timescales. Then, II) we may wish for a high temporal frequency for our forecasts. This then sets the upper bound on the model timestep to the desired frequency unless a second step is carried out where we do interpolation. If interpolation is used, we are further constrained by III) the number of timesteps the interpolation model can reliably bridge, which again imposes an upper bound. Finally, IV) distributions for a larger timestep will tend to be more multimodal and complex than for a smaller timestep, and may require more compute/larger models than you have available. Although these distributions will eventually simplify as they converge to climatology, this would no longer useful. 

We note that even with an interpolation model, the chosen model timestep will never exceed the timescale over which predictability is lost. This is because loss in predictability sets an upper bound on the interval an interpolator can meaningfully bridge. For example, imagine we attempt to interpolate between $X_0$ and a forecasted state $X_{1000}$, where $X_{1000}$ is so far in the future that it is essentially independent of $X_0$ i.e. memory of initial conditions is lost. The forecast of $X_{1000}$ would essentially be a sample of the climatology, and the interpolation model gains no useful information from it. In this case, interpolation would be no better than a model predicting purely from $X_0$. Thus, the maximum timestep that an interpolation model can effectively bridge must be shorter than the timescale over which predictability is lost. 

We therefore are left to answer the question: how large can the model timestep be before hitting one of the limits of
\begin{enumerate}
    \item the limit of predictability,
    \item the desired temporal frequency (if no interpolation is used),
    \item the interpolation model's capability (if interpolation is used), which must be shorter than the predictability limit, or
    \item degradation of results due to model capacity limits.
\end{enumerate}

As with typical ML regularization tuning, we monitor train and validation metrics. If the validation performance is worse than the training set, increasing regularization (e.g., using a longer timestep) is beneficial. If train and validation performances are similar, further regularization may be unnecessary and may actually degrade results without a corresponding increase in model capacity. We propose tracking two metrics in particular: forecast skill (e.g. CRPS as in Figure \ref{fig:model_timestep_comparison}a)) on the validation set and the forecast skill ratio (validation skill relative to training skill, as in Figure \ref{fig:l96_regularization}c)). To determine the model timestep, first, establish the upper bound: the maximum of I) the limit of predictability, II) the required resolution (if no interpolation is used) or III) the maximum timestep the interpolation model can handle. Train an initial model (e.g., with a 1-day timestep). If the forecast skill ratio indicates overfitting (validation skill significantly worse than training skill), increase the timestep. Check absolute forecast skills to confirm improvement, ensuring the model is not simply performing poorly on both train and validation sets. If overfitting persists (skill ratio $> 1$), continue increasing the timestep until either performance degrades (indicating model capacity limits, as seen for the 84-day model in Figure \ref{fig:model_timestep_comparison}c) and d)) or the upper bound is reached. If the skill ratio reaches 1, overfitting is no longer a concern, and further improvements should focus on increasing model capacity rather than regularization.

\section{Benchmarking on the ORAS5 dataset}
\label{section:seas5_experiments}

The previous sections showed that using a longer model timestep can improve results due to better regularization. This section is an illustration to show that longer-than-standard timestep models do also create realistic forecasts when compared to existing baselines. To do this, we compare SeaDiffusion with a 28-day timestep to the state-of-the-art SEAS5 \citep{johnson2019seas5}, persistence and climatology. SEAS5 contains 51 ensemble members, so we used 51-member ensembles of our model for all evaluation. Our version with a 28-day timestep generates a 196-day forecast in 50 seconds on a NVIDIA V100 device, and an ensemble of forecasts can be generated in parallel.

Our 28-day step is longer than those currently used in work using ML to model the ocean. For example, Ola \citep{wang2024coupled} uses a 48-hour step and DL\textit{ESy}M \citep{cresswell2024deep} use a 4-day step. ClimaX \citep{nguyen2023climax} makes predictions of particular variables upto 1 month ahead, but these are not able to be used to make a subsequent autoregressive forecast. \citet{arcomano2023hybrid} use a 7 day timestep for their SST model. 

In the following exercise, we evaluate against ORAS5. During our 2019-2022 test period, we initialize our model using ORAS5 at the first day of each month. We follow standard verification practice in evaluating ensemble forecasts using the single deterministic ORAS5 reanalysis as ground truth. Of the variables which our model tracks, we only compare sea surface temperature with SEAS5 because it is the only SEAS5 ocean variable publicly available via the Copernicus Climate Change Service (C3S) \citep{copernicus2018seasonal} on a daily resolution, since the rest of the variables have not yet been verified. We bias correct both SeaDiffusion and SEAS5 forecasts by subtracting a lead-time and calendar month specific mean bias, which is calculated based on the difference between model reforecasts and ORAS5 for 1993-2015. We regrid SEAS5 to a $1.5\degree$ grid from the $1\degree$ grid which it is provided on at C3S. SEAS5's native resolution is $0.25\degree$, and the ECMWF regrid it to $1\degree$ grid for C3S.

\subsection{Realism of samples}

\begin{figure}[h!]
\centering
\includegraphics[width=0.9\linewidth]{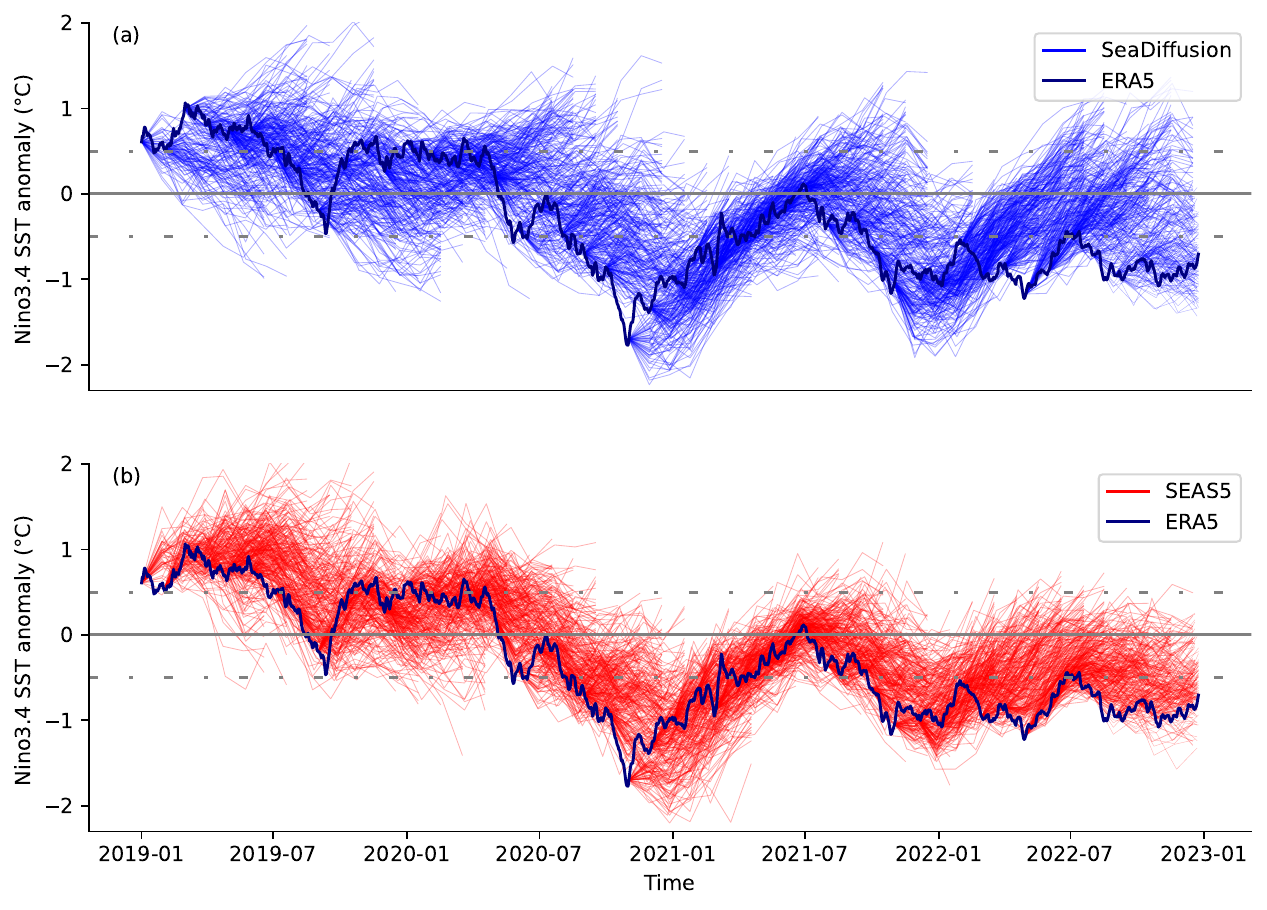}
\caption{Niño3.4 sea surface temperature anomaly plumes. Panels show 51-member ensemble forecasts from (a) SeaDiffusion (green lines) and (b) SEAS5 (red lines), initialized on the first day of each month from January 2019 to October 2022, with forecasts extending 196 days ahead. Anomalies are calculated relative to the ERA5 1981–2010 climatology. The ERA5 reanalysis, representing the ground truth, is highlighted in bold blue.}
\label{fig:sst_plumes_new}
\end{figure}

Figure \ref{fig:sst_plumes_new} illustrates SeaDiffusion's samples of Niño3.4 sea surface temperature (SST) anomaly plumes spanning from 2019 to 2023. SST anomalies in this region are critical indicators of ENSO phases, which significantly influence global weather patterns, including precipitation distribution, hurricane activity, and temperature fluctuations. 

Anomaly plumes are generated by averaging the SST over the Niño3.4 region and comparing these averages to the long-term climatology SST for that area (in this case we use ERA5 1981-2010 climatology as is standard). This process highlights deviations from typical temperature conditions, visualizing periods of anomalously warm or cool sea surface temperatures that correspond to El Niño or La Niña events, respectively. Specifically, El Niño events are typically characterized by SST anomalies exceeding +0.5°C, while La Niña events are identified by anomalies below -0.5°C. 

In Figure \ref{fig:sst_plumes_new}a, SeaDiffusion's generated plumes are compared with those produced by  SEAS5, displayed in Figure \ref{fig:sst_plumes_new}b. Qualitatively, SeaDiffusion demonstrates the ability to generate realistic SST anomaly trajectories that closely mirror SEAS5's outputs up until mid-2021. However, following mid-2021, SeaDiffusion encounters difficulties in accurately representing the latter stages of the unprecedented triple-dip La Niña event --- the first such occurrence in the 21st century. This event is characterized by three consecutive years of La Niña conditions, and presented significant challenges for predictive models. Although SEAS5 performs better, it also struggles in capturing the triple-dip event.

\subsection{Ensemble skill}

\begin{figure}[h!]
\centering
\includegraphics[width=1\linewidth]{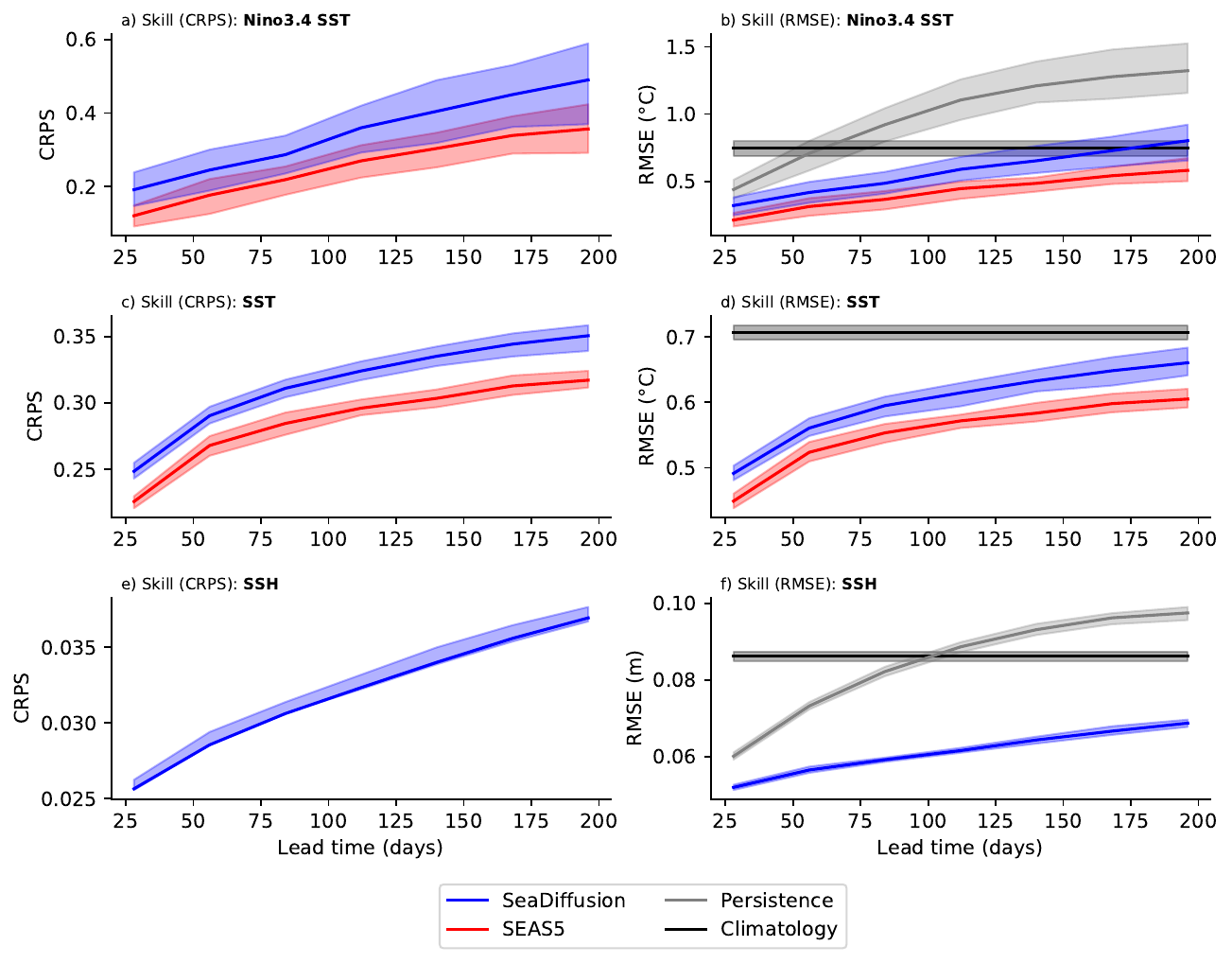}
\caption{Skill scores for 2019-2022. This compares the skill (y-axis) in terms of CRPS ((a), (c), (e)) and RMSE ((b), (d), (f)) of SeaDiffusion (green lines), SEAS5 (red lines), Persistence (grey lines) and Climatology (black lines), as a function of lead time (x-axis), for Nino3.4 sea surface temperature (Nino3.4 SST), whole-ocean sea surface temperature (SST), and whole-ocean sea surface height (SSH).  In all cases, lower is better. Error bars represent $95\%$ confidence intervals. Calculating CRPS require ensembles, which is not possible for Persistence and Climatology, hence they do not feature in the CRPS plots. Persistence is not shown in (d) because it is worse than Climatology for all lead times. SEAS5 is not present in (e) and (f) because SEAS5 SSH data is not publicly available.}
\label{fig:ensemble_skill}
\end{figure}

To evaluate SeaDiffusion's performance compared to SEAS5, persistence and climatology, we report CRPS and RMSE for Nino3.4 SST, SST and SSH. As shown in Figure \ref{fig:ensemble_skill}, SeaDiffusion is better or comparable with persistence and climatology in all cases. It is worse than SEAS5 for global SST, and for Nino3.4 SST the confidence bands overlap. 

\subsection{Ensemble calibration}


A well-calibrated probabilistic forecast should be confident when correct and uncertain when incorrect. In the weather community, this is evaluated using spread/skill ratios and rank histograms. Ideally, the members of an ensemble forecast should be indistinguishable from ground truth values, a property assessed via rank histograms. A flat rank histogram indicates that the truth is consistently indistinguishable from the ensemble members, an n-shaped histogram suggests over-dispersion with the truth ranking near the center, and a u-shaped histogram implies under-dispersion with the truth ranking near the tails.

\begin{figure}[h!]
\centering
\includegraphics[width=1\linewidth]{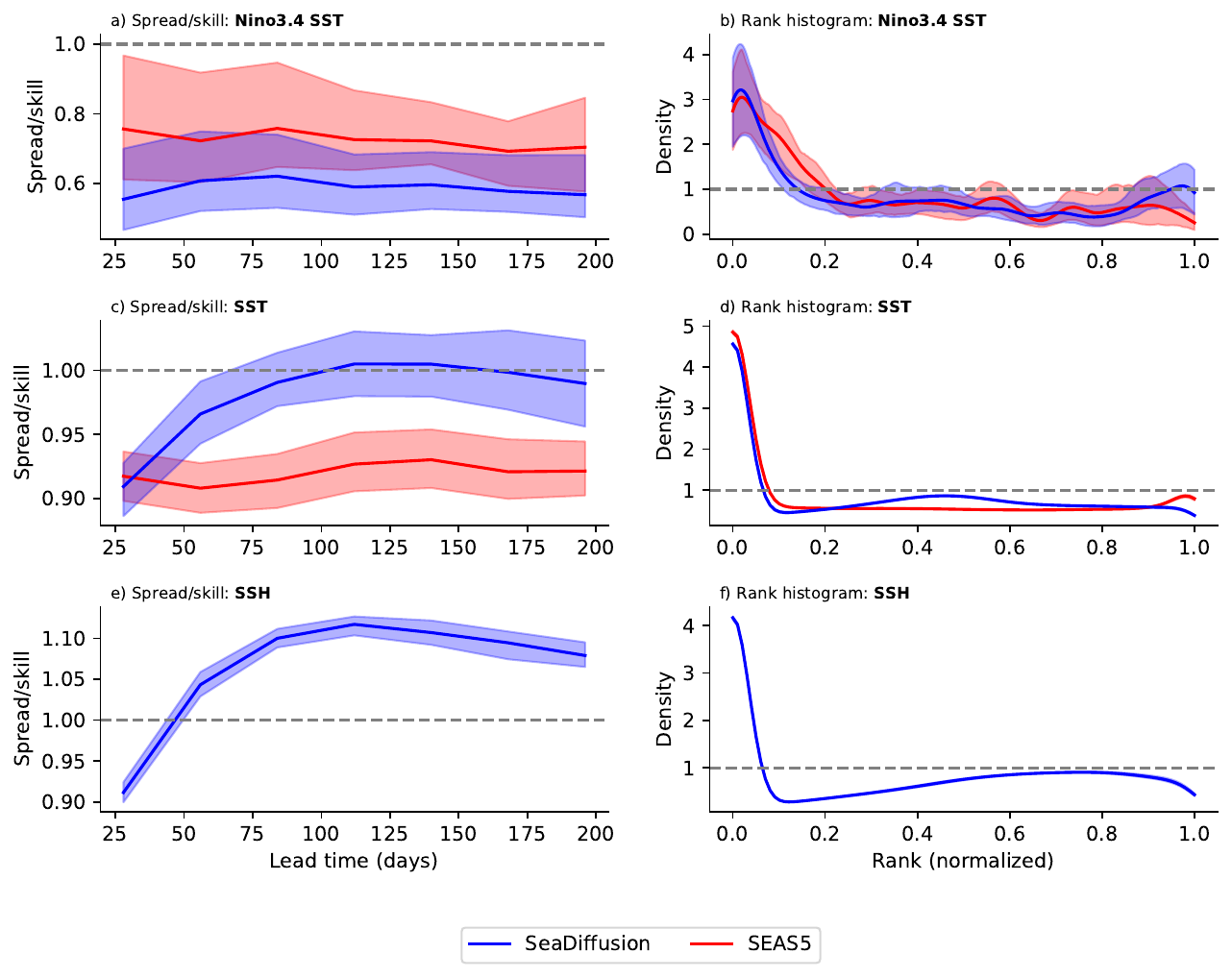}
\caption{Spread/skill scores and rank histograms for 2019-2022. This compares the skill (y-axis) in terms of CRPS ((a), (c), (e)) and RMSE ((b), (d), (f)) of SeaDiffusion (green lines), SEAS5 (red lines), Persistence (grey lines) and Climatology (black lines), as a function of lead time (x-axis), for Nino3.4 sea surface temperature (Nino3.4 SST), whole-ocean sea surface temperature (SST), and whole-ocean sea surface height (SSH).  In all cases, lower is better. Error bars represent $95\%$ confidence intervals. Calculating CRPS require ensembles, which is not possible for Persistence and Climatology, hence they do not feature in the CRPS plots. Persistence is not shown in (d) because it is worse than Climatology for all lead times. SEAS5 is not present in (e) and (f) because SEAS5 SSH data is not publicly available.}
\label{fig:ensemble_calibration}
\end{figure}

Figure \ref{fig:ensemble_calibration} shows spread/skill and rank histograms. In terms of spread/skill, SeaDiffusion's calibration is close to 1 for both SST and SSH, and is better than SEAS5 for SST. For SSH it is slightly overdispersive. For Nino3.4 SST, SeaDiffusion is undispersive and worse than SEAS5, although the confidence bands do overlap. In terms of rank histograms, SeaDiffusion performs similarly to SEAS5 for Nino3.4 SST and global SST. In all the rank histogram plots the peak at the smaller ranks occurs because both SEAS5 and SeaDiffusion overpredict the SST during the triple-dip La Nina event from mid-2021 to 2023.

\section{Future work}

\subsection{Regularization from coarser spatial resolution}

A coarser spatial resolution may also provide similar regularization by creating more cases of SIDO training pairs. Operationally, a forecast could be made at a coarse spatial resolution, and then upsampled to the desired resolution. Based on the Aurora model \citep{bodnar2024aurora}, it seems that a $0.25\degree$ spatial resolution for the backbone of models may be sufficient for weather and atmospheric forecasting cases. 

\subsection{Interpolating to increase temporal resolution}

Using a longer timestep for a forecast model yields a forecast with coarser temporal resolution, but this does not mean we must only use this model for our forecasts. Once this forecast has been made, an interpolation step can take place to `fill in the gaps'. We suspect this will be an easier task than the forecasting task, which is more akin to extrapolation. 

The interpolation model could either be a separate model to the forecaster or the same model if trained appropriately. A separate interpolator could take in states $X_t$ and $X_{t+\Delta t}$ and return inbetween states at a specified resolution. A combined forecaster-interpolater could be trained using a shuffled training objective \citep[e.g.][]{lessig2023atmorep,nguyen2022transformer,nivron2023taylorformer} so it learns to forecast arbitrary points in a sequence given an arbitrary conditional set. 

\subsection{Applicability to other tasks}

We suspect there are many other tasks where using a longer timestep naturally leads to SIDO regularization. Tasks across the Earth system are an obvious example, such as weather forecasting and subseasonal-to-seasonal forecasting. With the Earth system being chaotic, we expect that as the timestep increases, we see more cases of SIDO. At a broader level, generative models with longer timesteps could aid in ML solvers for ODEs, leading to quicker and perhaps more accurate solutions. 

\subsection{What is the use of the atmosphere when using ML for long-range forecasting?}

Oceanic initial conditions are the primary source of predictability for long-range forecasting, with the memory of the atmospheric initial conditions being lost after around two weeks. Given this, if we use a timestep greater than two weeks what input atmospheric variables are even useful when modelling the ocean's next state? Or are they all just forgotten? With generative modelling allowing us to model larger timesteps, the atmospheric variables could be treated as diagnostic instead of prognostic variables. A long-range forecast is then made by first evolving a minimal set of variables through time, and then snapshots of the corresponding atmosphere are generated based on this underlying forecast. 

\subsection{Architecture}

The ORAS5 experiments presented in this paper used a version of the conditional diffusion model ADM \citep{karras2024analyzing} with minor modifications. More sophisticated architectures tailored to the ocean might work better. For example, ingesting a land-sea mask as input, or using partial convolutions \citep{liu2018image} might help deal with land-sea artefacts in the current simulations. Conditioning on further past states, such as in GenCast \citep{price2025probabilistic}, may also help.

\section{Conclusions}

This paper shows that using ML models with a larger timestep provides regularization when forecasting the Earth system. We show the benefits of a longer timestep. Through careful experimentation, we show that improved regularization is one of the reasons for this. We also show how the regularization mechanism is due to more cases of similar input-different output in the training dataset, which is a well-established way to regularize models. We then provide a procedure for picking the model timestep to use. We also present a benchmarking exercise on the ocean reanalysis dataset, ORAS5, to demonstrate that models with longer-than-standard timesteps create realistic results. We suggest that there are many more applications of this regularization within the Earth system domain.





\subsubsection*{Author contributions}

Conceptualization, R.P.; Methodology, R.P. and M.A.; Software; R.P. and M.A.; Investigation, R.P. and M.A.; Writing -- Original Draft, R.P.; Writing -- Review \& Editing, R.P., M.A., H.M.C., F.V., D.J.W., J.Z.; Supervision, F.V. and D.J.W.

\bibliography{sample}

\clearpage

\appendix





\section{Datasets}

\subsection{Lorenz 96}
\label{appendix:l96}

For our experiments, we used the configuration of the Lorenz 96 as described in \citet{gan_hannah}. Using a value of $20$ for the `forcing' parameter, we created a dataset of length 5000 model time units (i.e. $1,000,000$ timesteps) and split it as follows. For the experiments in Section \ref{section:results}, we used the first 60\% for training, the penultimate 10\% for validation and the last 10\% for testing. For the experiments in Section \ref{section:benefits_regularization}, we created a test set for all data where either $X_t^0$ or $X_t^1$ or $X_t^2 > 14$. The validation set was for data where $X_t^3 > 14$, and the training set was the rest of the data. This split was done to help test generalizability to new input data.

\subsection{ORAS5}
\label{appendix:oras5}

We trained and evaluated our model on ECMWF's ORAS5 \citep{zuo2019ecmwf}, which is a global ocean reanalysis product. It represents the global ocean and is available from 1993 to present at a $0.25 \degree$ latitude/longitude resolution, and 1 day increments, for various variables at 75 depth levels. This covers the `altimetric era', where altimeter altimetry data observations became available after 1993. ORAS5 uses the same ocean model and sea-ice as the forecasts in SEAS5, and is driven by ocean observations from floats, buoys, satellites and ships. 

Our ORAS5 dataset contains a subset of the available ORAS5 variables (Table \ref{tab:variables_used}), on 6 depth levels: 0.5, 9.8, 47.2, 97.0, 199.8 and 300.9 m. 

We trained our model on 25 years of ORAS5 from 1993 to 2017, using 2018 as a validation year. Testing was done on 2019-2022.

\begin{table}[]
\centering
\begin{tabular}{@{}lll@{}}
\toprule
Type    & Variable name                   & Role            \\ \midrule
Oceanic & Sea water potential temperature & Input/Predicted \\
Single  & Sea surface height              & Input/Predicted \\ \midrule
Static  & Latitude                        & Input           \\
Static  & Longitude                       & Input           \\
Clock   & Elapsed year progress           & Input           \\ \bottomrule
\end{tabular}
\caption{ORAS5 variables used in our model. The `Type' column indicates whether the variable represents a time-varying oceanic property, a time-varying single-level property, a static property, or a property related to time itself. The `Role' column indicates whether the variable is something our model takes in as input and predicts, or only uses as input.}
\label{tab:variables_used}

\end{table}





\section{Models}

\subsection{Lorenz 96}

We trained our models with a batch size of 128, for 50,000 iterations. The models which performed best on the validation set were saved. We used Adam \citep{kingma2014adam} with a learning rate of 0.001. 

\subsection{SeaDiffusion}
\label{appendix:seadiffusion}

As in previous work on diffusion for Earth systems \citep{price2023gencast}, we model the residual between successive states, $p(X_{t+\Delta t} - X_t | X_t)$. We train separate models for different $\Delta t$. We broadly follow the framework presented by \citet{karras2022elucidating} and use the architecture presented in \citet{karras2024analyzing}. This architecture takes the ADM diffusion architecture \citep{dhariwal2021diffusion}, and improves its training dynamics by making modifications to preserve activation, weight and update magnitudes on expectation. 

Given the architecture involves downsampling and upsampling in the U-Net, we slice the 121 x 240 ORAS5 data to 112 x 240 to allow for easy down and upsampling. We do this by simply ignoring the lowest latitude levels. We use a maximum channel size of 64 for the convolution layers, which we do for computational reasons --- this tends to be far higher in industrial models, for example,  \citet{karras2024analyzing} use a maximum channel size of 768. We use a dropout rate of 0.3, and have 4 encoder/decoder blocks per downsampling/upsampling stage. We also modify the loss function from \citet{karras2022elucidating} to weight different grid cells based on their area, and to mask out values over land. 


All models were trained for 200,000 iterations with a batch size of 16 using Adam. The learning rate was decayed using a cosine scheduler from 5e-3 to 1e-5. Model hyperparameters were selected based on performance on the 2018 validation set. Our models were of size 20 million parameters. Training took 24 hours on a single V100 NVIDIA GPU.

\section{Further evaluation details}
\label{appendix:evaluation_details}

We use the same notation as in GenCast \citep{price2023gencast} to detail the evaluation metrics. In the following, for a particular variable, level and lead time,

\begin{itemize}
    \item For the SST and SSH variables, $x_{i,k}^m$ is the value of the $m$th of $M$ ensemble members in a forecast from initialization time indexed by $ k = 1,...,K$, at latitude and longitude indexed by $i \in G$. For the Nino3.4 SST and the Lorenz cases, we just have $x_k^m$. 
    \item For the SST and SSH variables, $y_{i,k}$ is the corresponding target. For the Nino3.4 SST and the Lorenz cases, we just have $y_k$. 
    \item For the SST and SSH variables, $\bar{x}_{i,k} = \frac{1}{M} \sum_m x_{i,k}^m$ is the ensemble mean. For the Nino3.4 SST and the Lorenz cases, $\bar{x}_{k} = \frac{1}{M} \sum_m x_{k}^m$.
    \item $a_i$ is the area of the latitude-longitude grid cell, which varies by latitude and is normalized to unit mean over the grid.
\end{itemize}

\subsection{CRPS}

In all cases, smaller is better. As in GenCast, we use the traditional CRPS estimator, which estimates the expected CRPS of the empirical distribution of a finite ensemble of size $M$. 

\subsubsection*{Lorenz 96 and Nino3.4 SST}

\begin{equation}
\operatorname{CRPS} \coloneqq \frac{1}{K}\sum_k \left( \frac{1}{M} \sum_m |x_k^m - y_k| - \frac{1}{2M^2}\sum_{m,m'} |x_k^m - x_k^{m'} |\right)
\end{equation}

\subsubsection*{Other ORAS5 variables}

\begin{equation}
\operatorname{CRPS} \coloneqq \frac{1}{K}\sum_k \frac{1}{|G|} \sum_i a_i\left( \frac{1}{M} \sum_m |x_{i,k}^m - y_{i,k}| - \frac{1}{2M^2}\sum_{m,m'} |x_{i,k}^m - x_{i,k}^{m'} |\right)
\end{equation}

\subsection{RMSE}

Again, smaller is better.

\subsubsection*{Lorenz 96 and Nino3.4 SST}

\begin{equation}
\operatorname{RMSE} \coloneqq \sqrt{\frac{1}{K}\sum_k  (y_{k} - \bar{x}_{k})^2}
\end{equation}

\subsubsection*{Other ORAS5 variables}

\begin{equation}
\operatorname{RMSE} \coloneqq \sqrt{\frac{1}{K}\sum_k \frac{1}{|G|} \sum_i a_i (y_{i,k} - \bar{x}_{i,k})^2}
\end{equation}

\subsection{Spread/skill ratio}

This should be close to 1 for a perfect forecast i.e. where the ensemble members and ground truth are exchangeable. 
For this, 
\begin{equation}
\operatorname{Skill} \coloneqq \operatorname{RMSE}
\end{equation}
and 
\begin{equation}
\operatorname{SpreadSkillRatio} \coloneqq \sqrt{\frac{M+1}{M}} \frac{\operatorname{Spread}}{\operatorname{Skill}},
\end{equation}
where spread is defined as follows:

\subsubsection*{Lorenz 96 and Nino3.4 SST}

\begin{equation}
\operatorname{Spread} \coloneqq \sqrt{\frac{1}{K}\sum_k \frac{1}{M-1} \sum_m \left(x_{k}^m - \bar{x}_{k}\right)^2}
\end{equation}

\subsubsection*{Other ORAS5 variables}

\begin{equation}
\operatorname{Spread} \coloneqq \sqrt{\frac{1}{K}\sum_k \frac{1}{|G|} \sum_i a_i \frac{1}{M-1} \sum_m \left(x_{i,k}^m - \bar{x}_{i,k}\right)^2}
\end{equation}

\section{Further results}

\begin{figure}[h!]
\centering
\includegraphics[width = 0.9\linewidth]{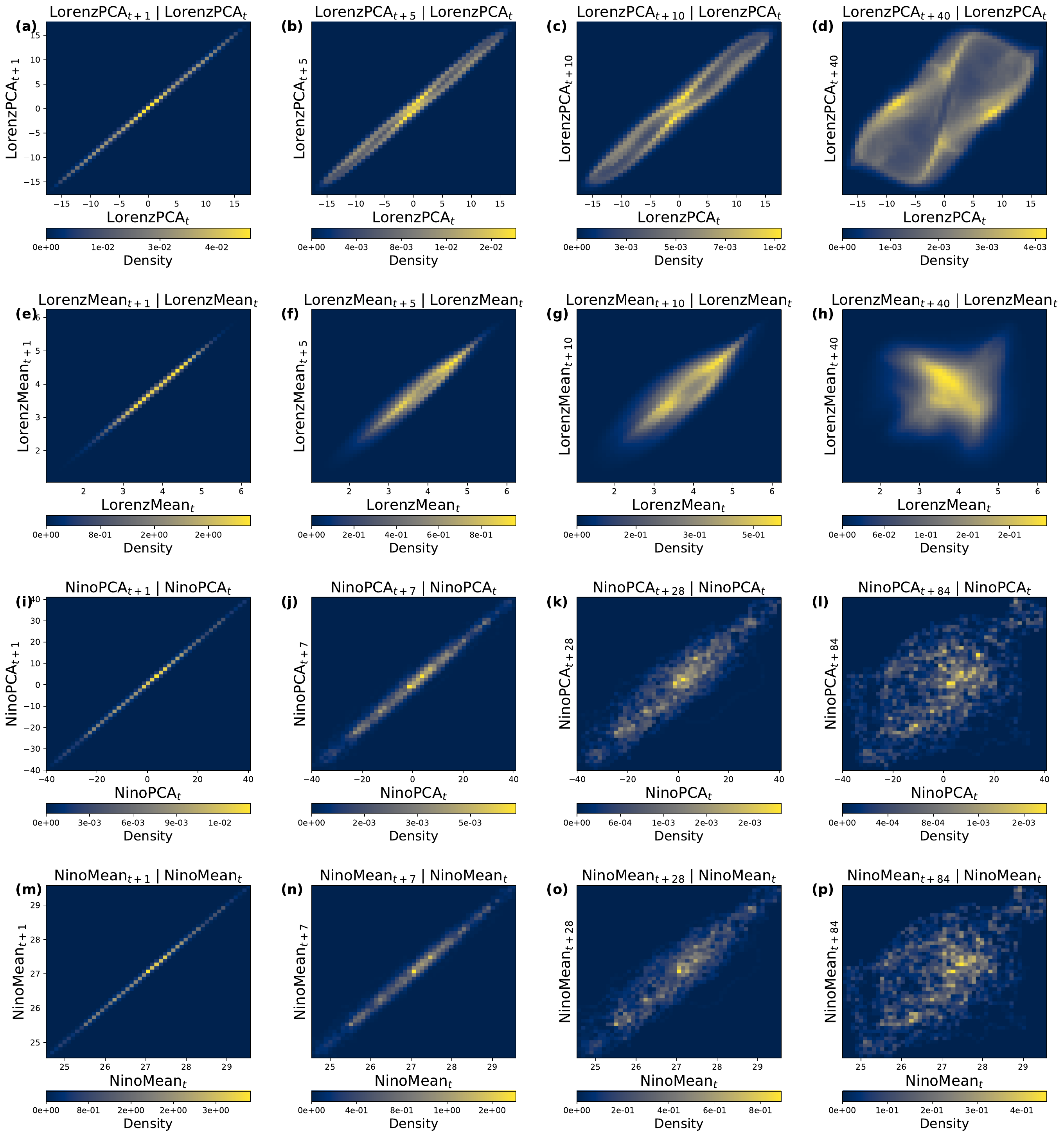}
\caption{2D histograms of PCA embeddings of Lorenz 96 (a, b, c, d), mean states of Lorenz 96 (e, f, g, h), PCA embeddings of the SST in the Nino3.4 region (i, j, k, l) and mean states of the SST in the Nino3.4 region (m, n, o, p). These histograms approximate the joint distributions. For the Lorenez 96 histograms, plots are shown for timesteps of timesteps of 1 x 0.005, 5 x 0.005, 10 x 0.005 and 40 x 0.005 model time units respectively. For the Nino histograms, plots are shown for timesteps of 1, 7, 28 and 84 days respectively.}
\label{fig:pca_mean_plots}
\end{figure}

\end{document}